\documentclass[reprint,superscriptaddress,amsmath,amssymb,aps,prl,]{revtex4-2}
\usepackage{ulem}
\usepackage{graphicx}
\usepackage{dcolumn}
\usepackage{bm}
\usepackage{physics}
\usepackage{amsmath}
\usepackage[colorlinks=true,citecolor=blue]{hyperref}
\usepackage{etoolbox}
\apptocmd{\sloppy}{\hbadness 10000\relax}{}{}
\usepackage{xcolor}
\usepackage[version=4]{mhchem}
\usepackage{bbm}
\usepackage{microtype}
\usepackage[capitalize]{cleveref}
\usepackage{ulem}
\usepackage{yfonts}
\usepackage[caption=false]{subfig}
\newcommand{\phantomsubfloat}[1]{
    {
        \captionsetup[subfigure]{labelformat=empty}
        \subfloat[][]{#1}
    }
}
\makeatletter
\newcommand{\overleftrightsmallarrow}{\mathpalette{\overarrowsmall@\leftrightarrowfill@}}
\newcommand{\overrightsmallarrow}{\mathpalette{\overarrowsmall@\rightarrowfill@}}
\newcommand{\overleftsmallarrow}{\mathpalette{\overarrowsmall@\leftarrowfill@}}
\newcommand{\overarrowsmall@}[3]{
  \vbox{
    \ialign{
      ##\crcr
      #1{\smaller@style{#2}}\crcr
      \noalign{\nointerlineskip}
      $\m@th\hfil#2#3\hfil$\crcr
    }
  }
}
\def\smaller@style#1{
  \ifx#1\displaystyle\scriptstyle\else
    \ifx#1\textstyle\scriptstyle\else
      \scriptscriptstyle
    \fi
  \fi
}
\makeatother

\begin{document}

\title{Proposal for spin superfluid quantum interference device}

\author{Yanyan Zhu}
\affiliation{Department of Physics and Astronomy and Bhaumik Institute for Theoretical Physics, University of California, Los Angeles, California 90095, USA}

\author{Eric Kleinherbers}
\affiliation{Department of Physics and Astronomy and Bhaumik Institute for Theoretical Physics, University of California, Los Angeles, California 90095, USA}

\author{Leonid Levitov}
\affiliation{Massachusetts Institute of Technology, Cambridge, Massachusetts 02139, USA}

\author{Yaroslav Tserkovnyak}
\affiliation{Department of Physics and Astronomy and Bhaumik Institute for Theoretical Physics, University of California, Los Angeles, California 90095, USA}

\begin{abstract}
In easy-plane magnets, the spin superfluid phase was predicted to facilitate coherent spin transport.
So far, experimental evidence remains elusive. 
In this Letter, we propose an indirect way to sense this effect via the spin superfluid quantum interference device (spin SQUID), inspired by its superconducting counterpart (rf SQUID). 
The spin SQUID is constructed as a quasi-one-dimensional (1D) magnetic ring with a single Josephson weak link, functioning as an isolated device with a microwave response. 
The spin current is controlled by an in-plane electric field through Dzyaloshinskii-Moriya interaction. This interaction can be interpreted as a gauge field that couples to the spin supercurrent through the Aharonov-Casher effect.
By investigating the static and dynamic properties of the device, we show that the spin current and the harmonic frequencies of the spin superfluid are periodic with respect to the accumulated Aharonov-Casher phase and are, therefore, sensitive to the radial electric flux through the ring in units of an electric flux quantum, suggesting a potential electric-field sensing functionality.
For readout, we propose to apply spectroscopic analysis to detect the frequency shift of the harmonic modes induced by this magnonic Stark effect.
\end{abstract}

\maketitle

\textit{Introduction.}\quad Spin superfluidity refers to the coherent spin transport that is mediated by topologically stable textures in easy-plane magnetic insulators~\cite{halperinPR1969}.
Analogous to conventional superfluids or superconductors, 
the spin superfluid is characterized by an approximate $U(1)$ order parameter describing the in-plane spin configuration~\cite{takeiPRL2014}.
If the order parameter winds up, the magnetic insulator sustains a nondissipative spin current~\cite{konigPRL2001,tserkovnyakPRL2017}.
In a ring structure, such a spin current is topologically protected and can unwind only via phase slips~\cite{soninAPhy2010}.
This effect can be found in both easy-plane ferromagnetic and antiferromagnetic systems~\cite{takeiPRL2014,takeiPRB2014}.

\begin{figure}[tbp!]
    \includegraphics[width=0.45\textwidth]{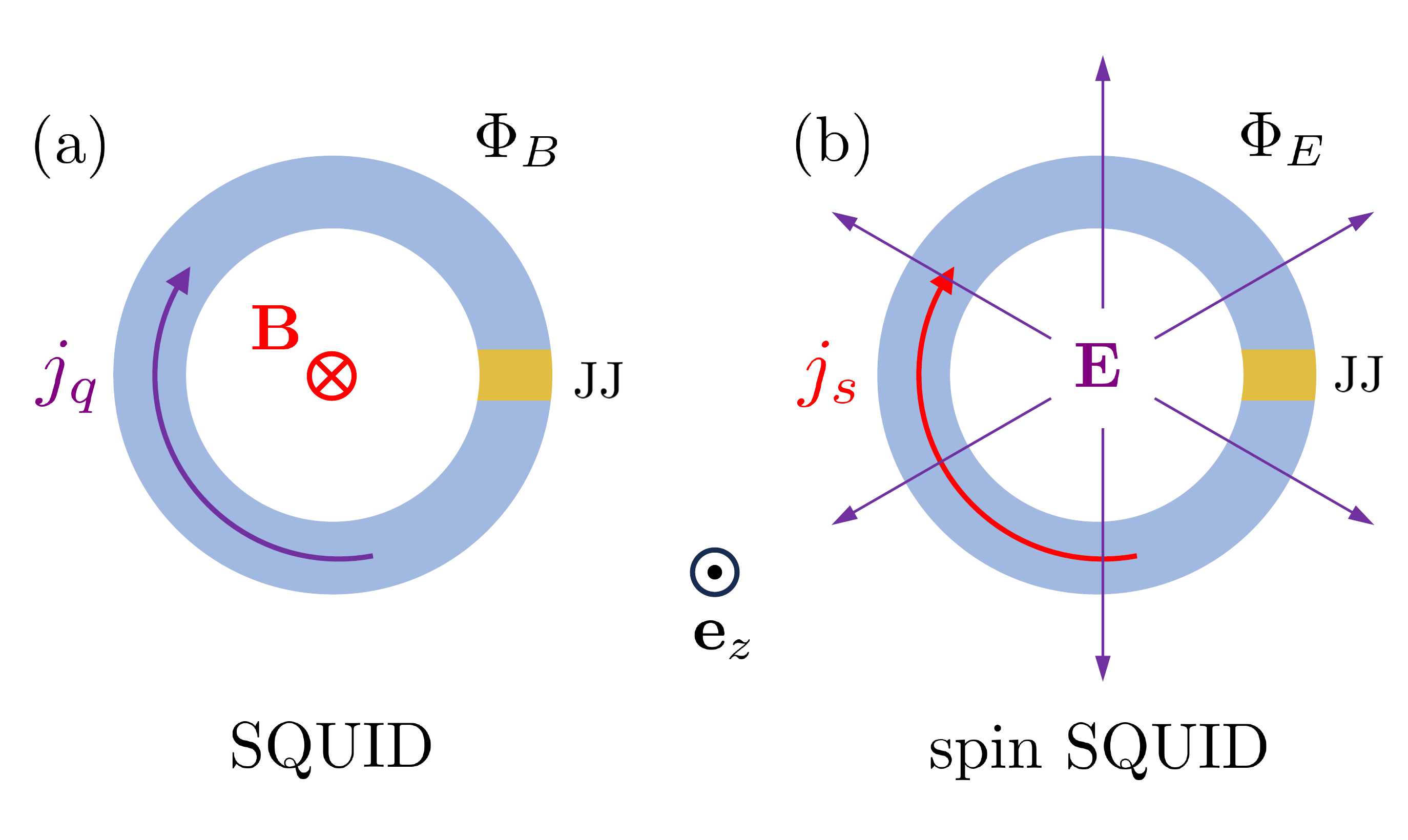}
    \phantomsubfloat{\label{fig: AB}}
    \phantomsubfloat{\label{fig: AC}}
    \vspace{-2\baselineskip}
    \caption{(a) A superconducting ring (blue part) with a Josephson junction (yellow part), where the charge supercurrent ${j}_q$ circulating in the ring is sensitive to the magnetic field $\vb{B}$ via the axial magnetic flux $\Phi_B =\int d S\, \vb{e}_z\cdot\vb{B}$ through the area enclosed by the ring.
    (b) A spin superfluid ring (blue part) with a weak-exchange link (yellow part), carrying a circulating spin current ${j}_s$ with spin in the $\vb{e}_z$ direction, is sensitive to the electric field $\vb{E}$ via the radial electric flux $\Phi_E=\oint d\ell\, \vb{e}_n\cdot\vb{E}$ through the ring.}
    \label{fig: AB AC}
\end{figure}

In a quantum-interference device, the topological nature of the superfluid phase winding in a ring structure is utilized to precisely measure externally applied fields that couple to the superflow.
For example, in the superfluid helium quantum interference device (SHeQUID), the phase of superfluid helium can sense the rotation of the earth~\cite{satoRPP2012}.
Another, more prominent example [see Fig.~\ref{fig: AB}] is the superconducting quantum interference device (SQUID), where the charge supercurrent of Cooper pairs accumulates the Aharonov-Bohm phase, which is sensitive to the magnetic flux $\Phi_B$ through the ring in units of the magnetic flux quantum~\cite{aharonovPR1959}.
In this Letter, we are interested in a dual version of this device [see Fig.~\ref{fig: AC}], termed the spin superfluid quantum interference device (spin SQUID), where the spin current accumulates the Aharonov-Casher phase, which is sensitive to the radial electric flux $\Phi_{E}$ through the ring in units of the electric flux quantum~\cite{aharonovPRL1984,meierPRL2003}.

For a basic spin SQUID, we propose a quasi-one-dimensional (quasi-1D) ring structure  with a single weak link, akin to the rf SQUID; see Fig.~\ref{fig: AC}. 
The ring (blue) is made of an easy-plane ferromagnetic insulator, and the weak link (yellow) with Heisenberg exchange coupling mimics a Josephson junction (JJ). 
Instead of reading out the low-frequency behavior of the adiabatic dynamics of the stationary state (as in the rf SQUID~\cite{clarkeWiley2004}), we propose to resonantly read out the first harmonic spin wave mode close a to a phase slip transition, which is nonlinearly shifted by the magnonic Stark effect. 
The spin SQUID is entirely based on the microwave response with no need for electric contact and any form of spin-to-charge conversion, 
which offers a simple and natural setting to confirm or even exploit the concept of spin superfluidity.
In practice, the spin SQUID could serve as an electric field detector, 
where the sensitivity can be tuned by the spin-orbit coupling strength.

\begin{figure}[htbp]
    \includegraphics[width=0.45\textwidth]{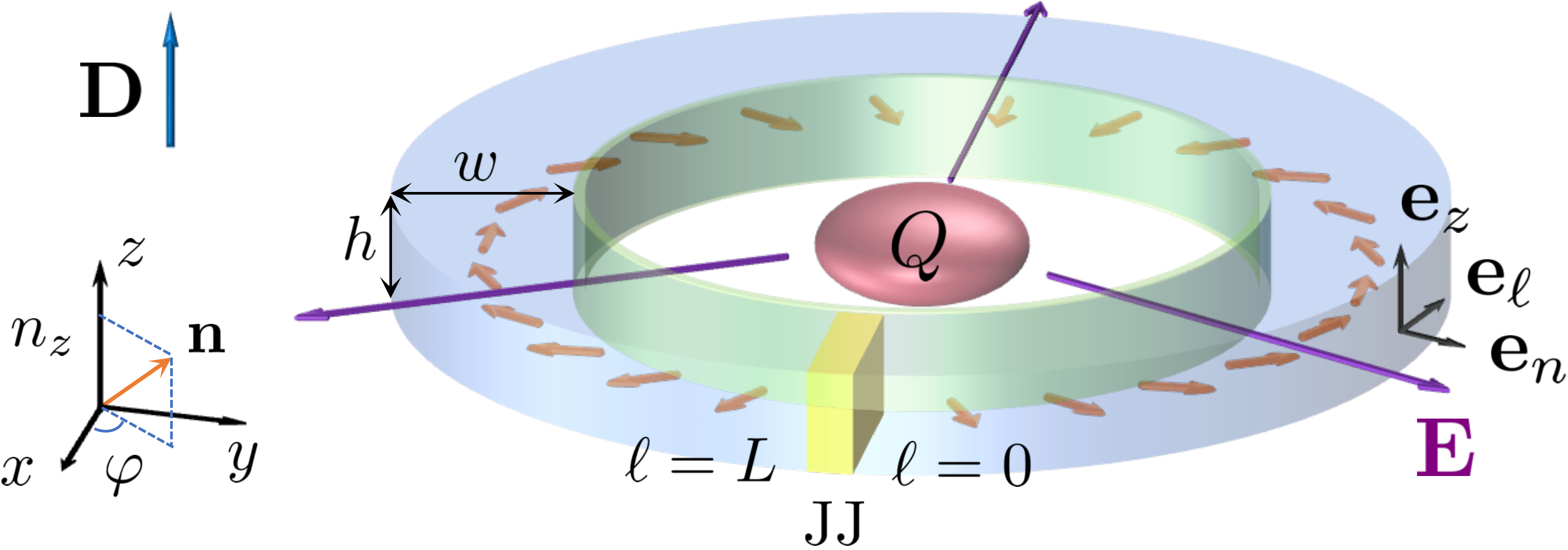}
    \caption{
    The blue part is the quasi-1D easy-plane ferromagnetic insulator with length $L$, width $w\ll L$, and height $h\ll L$. 
    The yellow part is a weak link modeled as a Josephson junction (JJ) and joins the 1D chain into a ring. 
    The electric charge $Q$ inside the ring produces in-plane electric fields at the radial direction $\vb{e}_n$. 
    The electric field induces an effective gauge field $a_E$ along the ring, which results in an equilibrium state with winding $\Delta\varphi$. 
    The green part is a heavy metal, which can enhance the DMI strength.}
    \label{fig: set up}
\end{figure}

\textit{Model.}\quad The spin SQUID shown in Fig.~\ref{fig: set up} is described by the following free energy: 
\begin{equation}
\begin{split}
    \label{eqn: free energy}
    \mathcal{F}=& \int_0^L d\ell\left[\frac{A}{2}(\partial_\ell\vb{n})^2+\frac{K}{2}n_z^2 -\vb{D}\cdot(\vb{n}\times\partial_\ell\vb{n})\right] \\
    &-F_\text{J}\vb{n}(0)\cdot\vb{n}(L), 
\end{split}
\end{equation}
where $\vb{n}$ is the unit vector along the local spin direction.
The first line describes the quasi-1D bulk of the ferromagnetic insulator
~\footnote{The system is required to be an insulator for two reasons: (1) to avoid Joule heating through itinerate electrons; (2) to avoid screening of the electric field which would suppress the DMI in our system.} 
of length $L$ with exchange stiffness $A$, easy-plane anisotropy $K$, and Dzyaloshinskii-Moriya interaction (DMI) $\vb{D}$.
The second line reflects the Heisenberg exchange coupling across the weak link, corresponding to an effective JJ with energy $F_\text{J}$. 
The coupling can be ferromagnetic ($F_\text{J}>0$) or antiferromagnetic ($F_\text{J}<0$), which mimics ordinary and $\pi$ Josephson junctions~\cite{normanAPL2024}, respectively. 
The DMI vector $\vb{D}=2\lambda A (\vb{E}\times\vb{e}_{\ell})$ is induced by the electric field~\cite{tserkovnyakPRB2007,tserkovnyakPRB2014,yangSCIR2018,srivastavaNANO2018,SM}, which locally breaks the inversion symmetry~\footnote{The DMI can be phenomenologically constructed from the microscopic Pauli spin-orbit coupling $H_R\sim\epsilon_{ijk}p_i\sigma_j E_k$ of the electrons, which linearly couples the spin current $\propto  p_i \sigma_j$ to the electric field $E_k$. This translates to a linear coupling of the spin current $ -A\vb{n}\times\partial_\ell\vb{n}$ to the electric field $\vb{E}$ in the magnetic free energy ${\cal F}$.}.
The strength of the effect is determined by $\lambda$, which is a phenomenological spin-orbit parameter that depends on materials.
In vacuum, $\lambda_{\mathrm{vac}}=e/4m_e c^2$, where $-e$ is the electron charge, $m_e$ is the electron mass, and $c$ is the speed of light. It can be several orders of magnitude larger in crystals~\cite{engelPRL2005}. 
We stress that the considerations below can be naturally generalized to an antiferromagnet, after replacing $n_z$ and the easy-plane anisotropy $K$ by the out-of-plane spin density and inverse spin susceptibility, respectively~\cite{takeiPRB2014}.

\textit{Equations of motion.}\quad We focus on a ferromagnet with strong easy-plane anisotropy, which possesses an approximate $U(1)$ symmetry~\cite{takeiPRL2014}.
In this case, we can parametrize $\vb{n}$ by its $z$ component $n_z$ and in-plane phase $\varphi$: $\vb{n}=(\sqrt{1-n_z^2}\cos{\varphi}, \sqrt{1-n_z^2}\sin{\varphi}, n_z)$.
The spin superfluid density corresponds to $1-n_z^2$.
In the strong easy-plane limit, where the length $L$ of the ring is much larger than the healing length $\lambda_{K}=\sqrt{A/K}$, we obtain, up to a constant~\cite{SM},
\begin{equation}
\begin{split}
    \label{eqn: free energy simplified}
    \mathcal{F}
    \approx&\int_0^L{d\ell\left[\frac{A}{2}(\partial_\ell\varphi-a_E)^2
    +\frac{\tilde{K}}{2}n_z^2\right]}   \\ 
    &-F_\text{J}\cos{(\varphi_L-\varphi_0)},
\end{split}
\end{equation}
where $\varphi_L\equiv\varphi(\ell=L)$ and $\varphi_0\equiv\varphi(\ell=0)$ are the in-plane angles at the weak link, and $a_E\equiv 2\lambda \vb{E}\cdot\vb{e}_{n}$ is an effective gauge field induced by the radial electric field via DMI.
Here, $\tilde{K} = K + Aa_E^2$ is the enhanced easy-plane anisotropy and the tilde will be omitted hereafter.
The out-of-plane electric field is of minor importance since out-of-plane spin winding is suppressed by the strong easy-plane anisotropy. 
In addition, we assume stability is guaranteed via the Landau criterion~\cite{soninAPhy2010}.
We also require a low temperature $T\ll K\lambda_K/k_\text{B}$ to keep the thermal fluctuations small $n_z\ll 1$.

The gauge field $a_E $ that 
couples to the in-plane phase winding resembles the magnetic vector potential that couples to the charge current.
We interpret the integrated gauge field as the Aharonov-Casher phase acquired when the magnetic moment is moving in an electric field.
The gauge field can also be spatially dependent, $a_E=a_E(\ell)$, which allows a deviation from cylindrical symmetry of the electric field $\vb{E}$ and the structure of the device.

Since $s n_z$ generates planar spin rotation, 
$(\varphi, s n_z)$ is a pair of conjugated variables~\cite{hillPRL2018} and satisfies the Poisson bracket 
$\{\varphi(\ell), s n_z(\ell')\}=\delta(\ell-\ell')$, where $s$ is the linear spin density.
The equations of motion can be derived from Hamilton's  equations~\cite{SM}:
\begin{align}
    s\partial_t n_z+\partial_\ell j_s&=-\alpha s \partial_t\varphi,    
    \label{eqn: EOMsnz} \\
    s\partial_t\varphi&=K n_z.    
    \label{eqn: EOMsphi}
\end{align}
Equation~\eqref{eqn: EOMsnz} is understood as the continuity equation for z spin, with the bulk spin current 
\begin{align}
j_s = -A(\partial_\ell\varphi-a_E),
\label{eqn: spin current}
\end{align}
and a loss term due to Gilbert damping described by the coefficient $\alpha$~\cite{tserkovnyakJAP2018}. 
Thus, the effective gauge field $a_E$ controls the spin current in the ferromagnet.
Equation~\eqref{eqn: EOMsphi} provides the analog to the ac Josephson relation, where $n_z$ plays the role of voltage~\cite{josephsonRMP1974}.

\begin{figure*}[tbp]
    \includegraphics[width=0.8\textwidth]{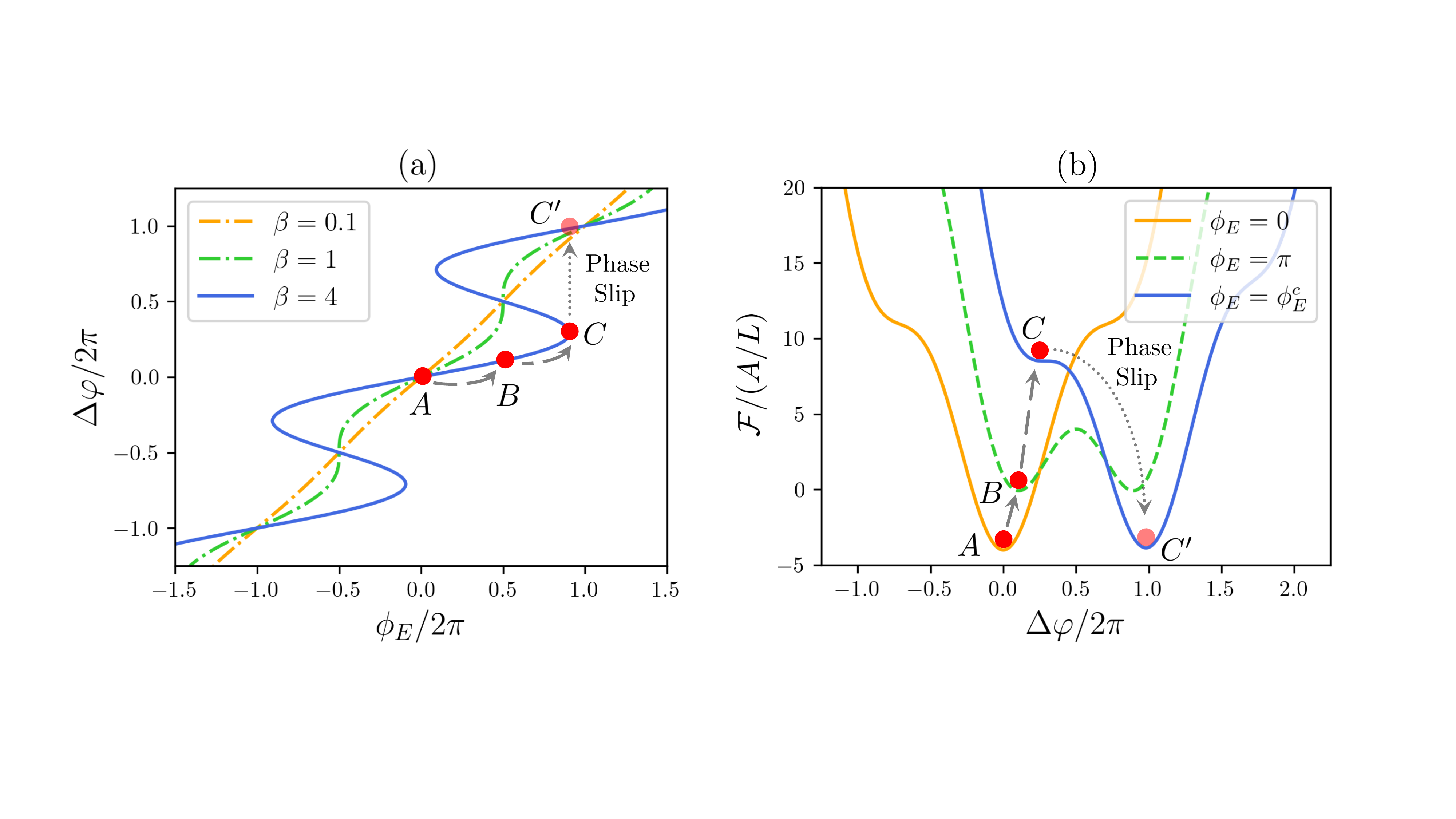}
    \phantomsubfloat{\label{fig: stationary condition}}
    \phantomsubfloat{\label{fig: free energy}}
    \vspace{-2\baselineskip}
    \caption{(a) Relationship between the winding $\Delta\varphi$ of stationary states and the accumulated phase $\phi_E$ for different $\beta$. 
    (b) Free energy ${\mathcal F}$ as a function of $\Delta\varphi$ for different values of $\phi_E$ when $\beta=4$. 
    In both (a) and (b), we indicate with red dots the ground state for $\phi_E=0$ (configuration $A$), the ground state for $\phi_E=\pi$ ($B$), beyond which the state becomes metastable, as well as the phase slip (from $C$ to $C'$), which is triggered at the critical flux $\phi_E=\phi_E^c$.}
    \label{fig: stationary states}
\end{figure*}

Boundary conditions can be established by enforcing the continuity of the spin current at the weak link. 
At the boundary, the spin current across the JJ is given by $j_\text{J}=F_\text{J} \sin{(\varphi_L-\varphi_0)}$.
Thus, the boundary condition becomes $j_s|_{\ell=0}=j_\text{J}$ and $j_s|_{\ell=L}=j_\text{J}$:
\begin{equation}
    \label{eqn: boundary condition}
    \begin{split}
    -A(\partial_\ell\varphi-a_E)|_{\ell=0}&=F_\text{J}\sin{(\varphi_L-\varphi_0)},  \\
    -A(\partial_\ell\varphi-a_E)|_{\ell=L}&=F_\text{J}\sin{(\varphi_L-\varphi_0)}.
    \end{split}
\end{equation} 

\textit{Equilibrium configuration.}\quad First, we calculate the stationary states of the model by minimizing the free energy~\eqref{eqn: free energy simplified}. We obtain $n_z=0$ and a constant spin current $j_s=-A\left(\partial_\ell\varphi-a_E\right)\equiv j_0$.
After integrating along the ring, we get $j_0 L=-A\left(\Delta\varphi-\phi_E\right)$, where we define the total winding $\Delta\varphi\equiv\int_0^L d\ell\,\partial_\ell\varphi = \varphi_L-\varphi_0$ and the accumulated Aharonov-Casher phase
\begin{align}
    \phi_E\ \equiv\int_0^L d\ell\, a_E = 
    2\pi\frac{\oint d \ell \,\vb{e}_n \cdot \vb{E} }{\Phi_E^0},
\end{align}
which is given by the dimensionless radial electric flux through the ring in units of the \textit{electric flux quantum} $\Phi_E^0=2\pi \hbar c/\tilde{g}\mu_\text{B}$ (in Gaussian units), where $\mu_\text{B}$ is the Bohr magneton and $\hbar$ is the reduced Planck constant~\cite{bogachekPRB1994,chenPRB2013,rostyslavPRB2023}. Here, $\tilde{g}\equiv 2\lambda\hbar c/\mu_\text{B}$ is a dimensionless factor enhanced by the spin-orbit coupling $\lambda$. In vacuum, it reduces to $\tilde{g}=1$.

For the stationary state, the boundary condition~\eqref{eqn: boundary condition} becomes:
\begin{equation}
    \label{eqn: stationary condition}
    \begin{split}
    0=&\ \Delta\varphi-\phi_E+\beta\sin{\Delta\varphi},
    \end{split}
\end{equation}
where the ratio $\beta\equiv \frac{F_\text{J} L}{A}$ is a crucial parameter that characterizes the nonlinear behavior of the spin SQUID. 
It can be estimated as $\beta \sim \frac{|\tilde{J}|}{J}\frac{L}{a}$, where $a$ is the lattice constant, $\tilde{J}$ is the exchange coupling at the JJ and $J$ is the bulk exchange coupling.
Despite the requirement $\tilde{J}\ll J$ for a nonlinear weak link, $\beta$ can be of $O(1)$ for systems with $L\gg a$.
$\beta$ also sets the amplitude of the spin current $\abs{j}_\text{max}= \beta A/L$. 
Then, the Landau criterion, $\abs{j}_\text{max}<\sqrt{KA}$, which ensures the stability of the spin superfluid becomes: $\lambda_{K}<L/\beta$~\cite{soninAPhy2010}.

The solution $\Delta\varphi(\phi_E)$ of Eq.~\eqref{eqn: stationary condition} is shown in Fig.~\ref{fig: stationary condition}.
For sufficiently weak coupling, $\abs{\beta}\ll 1$, the total winding $\Delta\varphi$ of the spin SQUID changes approximately linearly with the electric flux $\phi_E$.
For $|\beta|>1$, the relationship between the winding $\Delta\varphi$ and the electric flux $\phi_E$ becomes both nonlinear and multivalued, which enables \textit{phase slips}. 
In Fig.~\ref{fig: free energy}, we visualize the evolution of a stationary state and the phase slip using the free energy ${\mathcal F}(\Delta\phi)$ for different values of the electric flux $\phi_E$, 
where configuration $A$ is the ground state when $\phi_E=0$, $B$ is one of the two ground states at $\phi_E=\pi$~\footnote{This can be seen from the fact that both Eq.~\eqref{eqn: free energy simplified} and Eq.~\eqref{eqn: stationary condition} are symmetric under the transformation $\Delta\varphi\rightarrow-\Delta\varphi+2\pi$}, which will become metastable if $\phi_E$ is further increased, and $C\rightarrow C'$ presents the phase slip from the saddle point to the ground state when $\phi_E$ reaches the critical flux $\phi_E^c$ determined via $\partial_{\phi_E} \Delta \varphi\rightarrow\infty$.

It is insightful to rewrite Eq.~\eqref{eqn: stationary condition} as ${L_\text{ind}} j_s=-\Delta \varphi+\phi_E$, where we introduced the effective inductance $L_\text{ind}$. 
This equation is formally identical to that for the rf SQUID~\cite{clarkeWiley2004}---see Fig.~\ref{fig: AB}---after replacing the electric by the magnetic flux and the spin by the charge current. 
For the rf SQUID, the effective inductance has two contributions, $L_\text{ind}=L_\text{k}+L_\text{g}$, where the kinetic inductance $L_\text{k}$ stores energy in the kinetic inertia of the supercurrent and the geometric inductance $L_\text{g}$ stores energy in the generated magnetic field~\cite{shimazuPC2004}. 
For the spin SQUID, we have considered so far  only the kinetic contribution $L_\text{k}=L/A$ that stores energy in the exchange interaction. 
However, we also expect an analogous geometric contribution $L_\text{g}$~ that stores energy in the electric field generated by the spin current. 
In fact, we find that the spin current can be related to an electric polarization, which generates an electric field. 
The ratio between the geometric inductance and the kinetic inductance can be approximated as $L_\text{g}/L_\text{k}\sim A_\text{B}\lambda^2 {h}/(w+h)$, where $A_\text{B}=A /wh$ is the bulk exchange stiffness, and $w$ and $h$ are the width and height of the ring; see Fig.~\ref{fig: set up}~\cite{SM}.

\begin{figure*}[tbp]
    \includegraphics[width=0.8\textwidth]{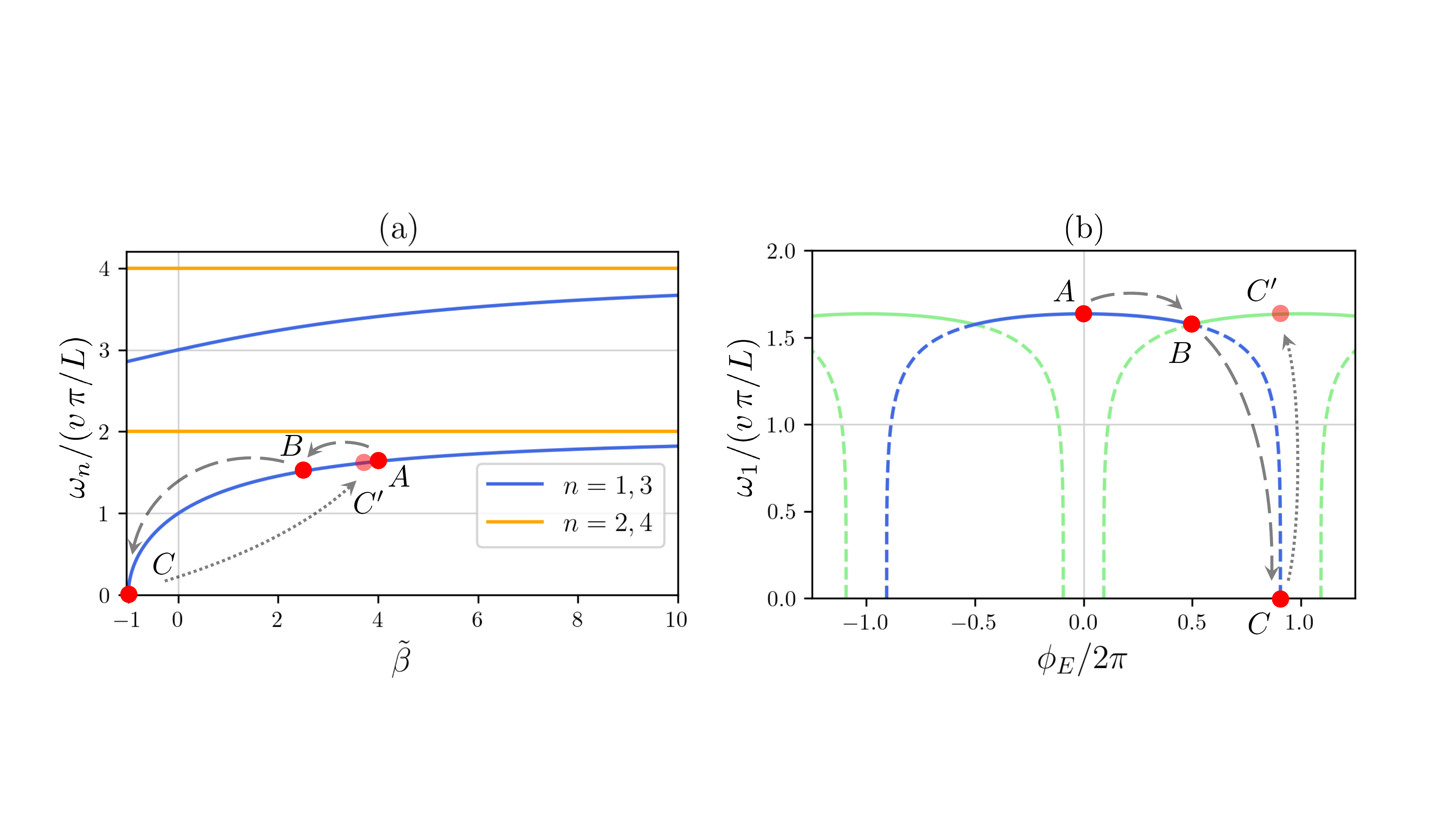}
    \phantomsubfloat{\label{fig: quantization condition}}
    \phantomsubfloat{\label{fig: k(phi)}}
    \vspace{-2\baselineskip}
    \caption{
    (a) Frequencies $\omega_n$ of the harmonic modes as a function of $\tilde{\beta}\equiv\beta\cos{[\Delta\varphi(\phi_E)]}> -1$. 
    Orange lines represent the even harmonic modes $\omega_{2m}$; and blue lines represent the odd harmonic modes $\omega_{2m-1}$ with $m\in{\mathbb{N}}$.
    (b) The first harmonic mode $\omega_1$ as a function of the electric flux $\phi_E$ at $\beta=4$.
    The solid lines represent the ground states and the dashed lines represent the metastable states.
    Configurations $A$, $B$, $C$ and $C'$ have the same meaning as in Fig.~\ref{fig: stationary states}, where $C\rightarrow C'$ is the phase slip and the region just before the phase slip is most sensitive to the flux change. 
    }
\end{figure*}

\textit{Dynamic properties.}\quad The nonlinearity of Eq.~\eqref{eqn: stationary condition} makes the spin SQUID sensitive to the electric flux $\phi_E$, which can be detected by the spin-wave modes on top of the wound-up ground state.
Since the stationary state has $n_z=0$ and $j_0=-A\left(\partial_\ell\varphi_\text{st} - a_E\right)$, the solution to the equations of motion~\eqref{eqn: EOMsnz} and \eqref{eqn: EOMsphi} can be expanded as: 
$\varphi=\varphi_\text{st}(\ell)+\delta\varphi(\ell,t)$, where $\varphi_\text{st}$ is the stationary solution, and $n_z=\delta n_z(\ell,t)$, where $\delta n_z\ll\delta\varphi\ll 1$. 
This leads to the wave equation for $\delta\varphi$ to linear order:
\begin{equation}
    \label{eqn: wave equation}
   \partial_t^2\delta\varphi+2\gamma \partial_t\delta\varphi-v^2 \partial_\ell^2\delta\varphi=0,
\end{equation}
where we defined the velocity $v=\sqrt{KA}/s$ and the dissipation rate $\gamma=\alpha K/2s$.
The resultant spectrum is $\omega=v k\sqrt{1-(\gamma/vk)^2}-i\gamma$.
For the first harmonic mode, the wavelength is comparable with the system length $L$, meaning $k_1 L\sim\pi$. 
We want the quality factor of the modes $Q\equiv \Re(\omega)/2 \Im(\omega)\gg 1$, so that the spin waves can communicate between the two ends of the spin SQUID before decaying.  
This requires $\gamma/v k_1 \sim \alpha L/\lambda_K\ll 1$, and the dispersion relation becomes: $\omega \simeq v k - i\gamma$. 
Therefore, the Gilbert damping imposes a restriction on the maximal value of the system size $L$. For example, with $\alpha\sim10^{-4}$~\cite{tremplerAPL2020} and $\lambda_K\sim 10\, a$~\cite{klyushinaPRB2017}, the system size $L/a$ should be much smaller than $10^5$.
We assume that the constraint is satisfied, and ignore the Gilbert damping $\alpha$ from now on.

The boundary condition of the wave equation~\eqref{eqn: wave equation} can be obtained from Eq.~\eqref{eqn: boundary condition}, by plugging in the 
ansatz of $\varphi$
\begin{equation}
    \label{eqn: boundary condition 02}
    \begin{split}
    -\left.L\partial_\ell\delta\varphi\right|_{\ell=0}&=\tilde{\beta}(\delta\varphi_L-\delta\varphi_0), \\
    -\left.L\partial_\ell\delta\varphi\right|_{\ell=L}&=\tilde{\beta}(\delta\varphi_L-\delta\varphi_0),
    \end{split}
\end{equation}
which describes the fluctuating spin current at the weak link.
Here, we define $\tilde{\beta}=\beta \cos{\Delta\varphi}$, which is bounded from below, $\tilde{\beta}\geq -1$, where $\tilde{\beta}=-1$ corresponds to the phase slip transition at $\phi_E^c$.
The resulting eigenmodes are even $\delta\varphi_\text{e} \propto \cos[k_\text{e}(\ell-\ell_0)]$ and odd $\delta\varphi_\text{o} \propto \sin[k_\text{o}(\ell-\ell_0)]$ with respect to the center of the ring $\ell_0=L/2$, where the wavenumbers $k_\text{e,o}\ge 0$ are quantized via
\begin{equation}
    \label{eqn: quantization condition}
    \tan{(k_\text{e}L/2)}=0  \quad\mathrm{and}\quad 
    \tilde{\beta}=\frac{-k_\text{o}L/2}{\tan{(k_\text{o}L/2)}}.
\end{equation}
In Fig.~\ref{fig: quantization condition}, we show the associated frequencies $\omega_{2m-1}=v k_\text{o}$ and  $\omega_{2m}=v k_\text{e}$ for $m\in{\mathbb{N}}$ in blue and orange. The frequencies of the even harmonic modes are given by
\begin{align}
    {\omega_{2m}}&= v \frac{2 m \pi}{L}.
\end{align}
Hence, the JJ is inoperative, because the motion of the spins at the weak link is always in phase. For odd harmonic modes, on the other hand, they are out of phase and, therefore, sensitive to the JJ.

The frequencies close to $\tilde{\beta}=0$ can be linearized as
\begin{align}
    \label{eqn:oddfrequency}
    {\omega_{2m-1}}&= v \frac{(2m-1) \pi}{L}  +\frac{4v\tilde{\beta}  }{(2m-1) \pi L} + O(\tilde{\beta}^2). 
\end{align}
For small coupling strength $\beta\rightarrow 0$, the weak link energy is negligible, recovering the result for open spin chains with frequencies $\omega_n\rightarrow v\,n\pi/L$. 
If the coupling is very strong $\beta\rightarrow\infty$, we get $\omega_n\rightarrow v\,2n\pi/L$ with two-fold degeneracy, consistent with the result for periodic spin chains. 
We refer to the dependence of the frequencies on the electric field as the \textit{magnonic Stark effect}.

We find that the frequency of the first harmonic mode $\omega_1$ is most sensitive to $\tilde{\beta}$; see Eq.~\eqref{eqn:oddfrequency} and Fig.~\ref{fig: quantization condition}. In fact, close to the phase slip at $\tilde{\beta}=-1$, we obtain
\begin{align}
    \omega_1\approx \frac{2 v}{L}\sqrt{3(1+\tilde{\beta})},
\end{align}
where the sensitivity $\partial_{\tilde{\beta}} \omega_1$ diverges as the phase slip is approached.
This can be used to sense the electric field. 
By inserting the stationary state $\Delta\varphi(\phi_E)$ from Eq.~\eqref{eqn: stationary condition}, we can obtain the relationship between the harmonic frequency $\omega_1$ and the electric flux $\phi_E$, as shown in Fig.~\ref{fig: k(phi)}. 
The highest sensitivity occurs when the stationary state is prepared in the metastable region (between points $B$ and $C$) and approaches the phase slip point ($C$). 
This can be done by cooling down the system to the ground state (point $A$) first and then adiabatically increasing $\phi_E$.
Additionally, we require a reasonably large $|\beta|>1$ to access the highly nonlinear regime, which is feasible in practice, as $\beta=\frac{\tilde{J}}{J}\frac{L}{a}$ can be of $O(1)$ by going to large L. 
Therefore, this first harmonic mode close to the phase slip transition ($\tilde{\beta}\rightarrow-1$) appears attractive for electric field sensing.
We remark that a finite temperature limits the sensitivity because the energy barrier of the metastable state may be overcome by thermal fluctuations as we approach the phase slip point.

\textit{Spectroscopic readout.}\quad In order to experimentally readout the frequency $\omega_1$ of the first harmonic mode, we propose to apply spectroscopic methods. 
For example, using ferromagnetic resonance (FMR) by applying a microwave field and scanning its frequency, one can measure the absorption spectrum of the spin SQUID at a certain electric flux $\phi_E$, and resolve the first harmonic mode $\omega_1$ from the resonant frequency~\cite{maksymovPE2015}. 
Then, we can obtain the corresponding electric flux $\phi_E$ from Fig.~\ref{fig: k(phi)} getting information on the local in-plane electric field as well as the charge distribution on submicron scale, as shown in Fig.~\ref{fig: set up}. 
The higher the $Q$ factor of $\omega$, the better the sensitivity is.
We may also place nitrogen-vacancy (NV) centers as a sensor to detect the harmonic modes via relaxometry~\cite{takeiPRR2024,casolaNATR2018}. 

\textit{Parasitic effects.}\quad Unlike the superfluidity of \ce{^4He} or an $s$-wave superconductor with exact $U(1)$ symmetry, the spin superfluid can be compromised by parasitic anisotropies or some external field, which break the $U(1)$ symmetry and result in phase fixation. 
We can estimate those parasitic effects as length scales, and focus on the case that the system size $L$ is sufficiently small in comparison.
For example, an in-plane anisotropy can cause phase fixation, where the spin in-plane angle does not wind up but forms domain walls. The size of the domain wall is the parasitic healing length $\lambda_{K'}=\sqrt{A/K'}$, where $K'$ parametrizes the strength of in-plane anisotropy~\cite{soninAPhy2010}, and we require $\lambda_{K'}\gg L$. 
In contrast, the beneficial length scales should be smaller or  at least the same order as $L$. For example, a smaller healing length $\lambda_K=\sqrt{A/K}$ could allow a larger winding to build up. 
One illustration of the parasitic effects is that a ferromagnet with saturation magnetization $M_s$ may suffer from the anisotropy inherent to its magnetostatic properties, with easy-plane anisotropy $K = 4\pi M_s^2$, and in-plane anisotropies $K'/K\sim h/w$, respectively. 
To minimize the impact of such a parasitic magnetostatic effect, we require the device to be in the ultrathin limit $h\ll w$, so that the constraint $\lambda_{K'}\gg L\gtrsim \lambda_K$ can be fulfilled.

\textit{Thermal gradient.}\quad Thermal gradients can provide an additional handle to generate and measure coherent phase winding. 
A thermal gradient applied to a superfluid conduit generally induces an entropy-carrying normal flow, which, in the steady state, would trigger a superfluid counterflow. 
The associated phase-gradient build-up can then shift the gauge-induced interference fringes. This has been utilized to construct a flux-locked SHeQUID interferometer~\cite{satoRPP2012}, which can be tuned to operate at an optimal rotation-independent sensitivity. 
We expect similar phenomenology in the case of the spin SQUID, where the thermally induced normal spin current polarized out of the easy-plane is related to the spin Seebeck effect~\cite{bauerNATM2012}. 
It is noteworthy that a normal magnetic field is needed here, in order to cant magnetic spins out of the easy-plane, allowing magnons to carry a net $z$ component of spin angular momentum. 
We can understand this according to symmetry requirements for a thermal gradient to induce a planar order-parameter winding (with the sign of the applied field controlling the sign of winding, for a given thermal gradient). 
The details of heating-based modalities for spin SQUID interferometry, which require two-fluid dynamical treatment, will be discussed elsewhere.

\textit{Conclusions.}\quad To exploit and thus demonstrate the existence of a spin superfluid, we propose the spin SQUID based on the analogy to the rf SQUID. 
By analyzing the free energy of the spin SQUID and coupling it to an effective gauge field that is linear in the in-plane electric field, 
we calculate the relationship between the stationary state and the electric flux thread radially through the device in units of the electric flux quantum.
Close to a phase slip, the frequency of the first harmonic spin wave mode is highly sensitive to the electric flux.
Thus, combined with spectroscopic methods, the spin SQUID can be a potential electric field detector at submicron scale.
We stress that the characteristic response of the spin SQUID to the electric flux is based on a \textit{nonlocal gauge effect}, as the accumulated Aharonov-Casher phase of the spin superfluid will only be effective when the weak link is turned on~\cite{krivoruchkoLTP2020}. 
Without the weak link, the harmonic frequency is independent of the electric flux. This would serve as an evidence for spin superfluidity.

\begin{acknowledgments}
\textit{Acknowledgments.}\quad This work is supported by the NSF under Grant No. DMR-2049979. 
Y.Z. acknowledges support from the Julian Schwinger Fellowship at UCLA.
\end{acknowledgments}

\normalem
\bibliography{bib_rf_V11}

\end{document}


\setcounter{equation}{0}
\setcounter{figure}{0}
\setcounter{table}{0}
\setcounter{page}{1}
\makeatletter
\renewcommand{\thefigure}{S\arabic{figure}}
\renewcommand{\bibnumfmt}[1]{[S#1]}
\renewcommand{\citenumfont}[1]{S#1}
\def\EK{\color{red}}

\title{Supplemental Materials for \texorpdfstring{\\}{} 
``Proposal for Spin Superfluid Quantum Interference Device"}

\author{Yanyan Zhu}
\affiliation{Department of Physics and Astronomy and Bhaumik Institute for Theoretical Physics, University of California, Los Angeles, California 90095, USA}

\author{Eric Kleinherbers}
\affiliation{Department of Physics and Astronomy and Bhaumik Institute for Theoretical Physics, University of California, Los Angeles, California 90095, USA}

\author{Leonid Levitov}
\affiliation{Massachusetts Institute of Technology, Cambridge, Massachusetts 02139, USA}

\author{Yaroslav Tserkovnyak}
\affiliation{Department of Physics and Astronomy and Bhaumik Institute for Theoretical Physics, University of California, Los Angeles, California 90095, USA}

\maketitle

\vspace{-2\baselineskip}


\section{Dzyaloshinskii-Moriya interaction} 
Dzyaloshinskii-Moriya interaction (DMI) arises from spin-orbit coupling 
which requires inversion symmetry breaking of the magnets~\cite{moriyaPR1960,dzyaloshinskyJPCS1958}. In the following, we derive the DMI~\cite{tserkovnyakPRB2014} which is phenomenologically given by 
\begin{equation}
    \label{eqn: DMI}
    \mathcal{F}_{\mathrm{DMI}}
    = -\int d\ell \,\vb{D} \cdot(\vb{n}\times\partial_\ell\vb{n}).
\end{equation} 
We start from a microscopic description of the electrons including spin-orbit interaction as described by the Pauli Hamiltonian~\cite{frohlichRMP1993}
\begin{align}
    {H}=\frac{\vb{p}^2}{2m_e}+\frac{\hbar \lambda}{2m_e} \left[\vb{p}\cdot \left(\boldsymbol{\sigma}\times\vb{E}\right)+ \left(\boldsymbol{\sigma}\times\vb{E}\right)\cdot \vb{p}\right]=\frac{1}{2m_e}\left(\vb{p}+ \hbar\lambda\boldsymbol{\sigma}\times\vb{E}\right)^2+{\cal O}(\lambda^2),
\end{align}
where the electric field $\vb{E}$, comprising both externally applied and intrinsic fields, corresponds to the local inversion symmetry breaking and can be $\ell$ dependent. $\vb*{\sigma}$ denotes the vector of Pauli matrices and $\lambda$ parameterizes the spin-orbit coupling which will reduce to $\lambda=e/4m_ec^2$ in vacuum with $-e$ and $m_e$ denoting the elementary charge and the mass of the electron, respectively~\footnote{This Hamiltonian can also be written as $H=\frac{1}{2m}\left(\vb{p} - \frac{\tilde{g}}{2 c}\vb*{\mu}\times\vb{E}\right)^2+{\cal O}(\lambda^2)$, where $\vb*{\mu}=-g_e\mu_\text{B}\hbar\vb*{\sigma}/2$ is the magnetic moment of the electron with g-factor $g_e$ and Bohr magneton $\mu_\text{B}=e\hbar/2m_e c$ and $\tilde{g}= 2\lambda\hbar c/\mu_\text{B}$ is a dimensionless factor which reduces to 1 in vacuum. The factor $1/2$ in front of $\vb*{\mu}\times\vb{E}$ comes from the Thomas precession and vanishes for charge-neutral particles~\cite{aharonovPRL1984,anandanPRA1989}}. As a first step, we apply the unitary transformation
\begin{align}
    {U}=\mathcal{{P}}_{\ell\leftarrow 0}\,e^{-i\lambda\int \limits_0^\ell d\ell^\prime \vb{e}_{\ell^\prime} \cdot\left[\vb*{\sigma}\times\vb{E}({\ell^\prime})\right]},
\end{align}
where $\mathcal{{P}}_{\ell\leftarrow 0}$ is the path ordering operator, which puts operators with larger positions $\ell$ to the left. Its effect is a  boost in momentum according to
\begin{align}
    {U}^\dagger (\vb{p}+\hbar \lambda\vb*{\sigma}\times\vb{E}) {U}= \vb{p}.
\end{align}
Thus, the Hamiltonian transforms to 
\begin{align}
    {\tilde{H}}= {U}^\dagger{H}{U}=\frac{\vb{p}^2}{2m_e},
\end{align}
which is isotropic. After adding electron-electron interactions, one can obtain an emergent low-energy theory for the  magnetic order parameter $\vb{\tilde{n}}$. The resultant exchange energy will be isotropic as well and takes the form 
\begin{align}
    {\mathcal F}_\text{X}=\int d\ell \,\frac{A}{2}\left(\partial_\ell\vb{\tilde{n}}\right)^2= 
    \int d\ell \,\frac{A}{2}\left({\cal R} \partial_\ell\vb{\tilde{n}}\right)^2. \label{eq:SMfree}
\end{align}
The relation from $\vb{\tilde{n}}$ to the order parameter $\vb{n}$ in the original space is given via $\vb{\tilde{n}}= {\cal R}^{-1} \vb{n}$, where the rotation matrix  ${\cal R}$ is the $\text{SO}_3$ representation of the unitary transformation ${U}$~\footnote{The  coupling between the electron spin and the magnetic order parameter transforms as $\vb*{\sigma}\cdot\vb{n}\rightarrow {U}^\dagger\vb*{\sigma}{U}\cdot\vb{n}={\cal R}\vb*{\sigma}\cdot\vb{n}=\vb*{\sigma}\cdot{\cal R}^{-1}\vb{n}$. Thus, the order parameter $\vb{n}$ transforms with the inverse transformation ${\cal R}^{-1}$~\cite{kimPRL2013}.}. In the second step, we inserted the same matrix ${\cal R}$ by using the invariance of the scalar product under rotations. 

The specific form of ${\cal R}$ can be obtained by viewing ${U}$ not as a boost in momentum, but as a rotation matrix for spinors with generator $\hbar \vb*{\sigma}/2$. Then, the  rotation matrix for vectors is obtained by replacing  the generator with $\hbar \vb{L}$, where $(L_i)_{jk}=-i \epsilon_{ijk}$. Thus, 
\begin{align}
    {\cal R}=\mathcal{{P}}_{\ell\leftarrow 0}\,e^{-i2\lambda\int \limits_0^\ell d\ell^\prime \vb{e}_{\ell^\prime} \cdot\left[\vb*{L}\times\vb{E}({\ell^\prime})\right]}.
\end{align}
The transformed derivative of Eq.~\eqref{eq:SMfree} becomes
\begin{align}
    {\mathcal R}\partial_\ell\vb{\tilde{n}}&=\partial_\ell \vb{n} +\left({\cal R} \partial_\ell {\cal R}^{-1}\right)\vb{n}\\
    &=\left[\partial_\ell - 2\lambda \left(\vb{E}\times\vb{e}_\ell\right)\times\right]\vb{n},
\end{align}
which takes the form of a chiral derivative~\cite{kimPRL2013}. 
Finally, the free energy can be evaluated as
\begin{align}
     {\mathcal F}_\text{X} =\int d\ell \,\frac{A}{2}\left[\partial_\ell\vb{{n}}-2\lambda \left(\vb{E}\times\vb{e}_\ell\right)\times\vb{n}\right]^2=\int d\ell \,\left[\frac{A}{2}\left(\partial_\ell\vb{{n}}\right)^2-2\lambda A\left(\vb{E}\times\vb{e}_\ell\right) \cdot \left(\vb{n}\times\partial_\ell \vb{n}\right) \right] + {\cal O}(\lambda^2),
\end{align}
where we found the DMI from Eq.~\eqref{eqn: DMI} by identifying $\vb{D}=2\lambda A\left(\vb{E}\times\vb{e}_\ell\right)$. 
This linear relation with the electric field is also illustrated in some experiments~\cite{srivastavaNANO2018,yangSCIR2018}. 
Although the above derivation is performed exactly in 1d, the result can be generalized to higher dimensions, with additional corrections of order $\lambda^2$.

For real materials, the bulk DMI strength can be around $D\sim 0.1\ \text{mJ/m}^2$~\cite{zhangPRL2022}. Assuming a bulk exchange stiffness $A\sim 4\times 10^{-12}\ \text{J/m}$~\cite{klinglerJPD2015}, we estimate the spin-orbit length $\ell_\text{SO}=2\pi A/D \sim 0.25\ \mu\text{m}$, which quantifies the length over which the spin winds by $2\pi$. 
According to Refs.~\cite{srivastavaNANO2018,yangSCIR2018}, the tunability of the DMI strength by the electric field can also be large, with the adjustment range of DMI to be about $\Delta D\sim 0.08\ \text{mJ/m}^2$. 
For the operation of the spin SQUID, these requirements, together with a strong easy-plane anisotropy and low Gilbert damping, are sufficient if they are achieved simultaneously in the same material.

In the main text, we consider the easy-plane limit, where the direction of the spin associated with the spin current points in the $z$ direction: $\vb{n}\times\partial_\ell \vb{n}\propto \vb{e}_z$. Thus, only the radial electric field $\vb{E}\propto \vb{e}_\ell\times\vb{e}_z$ couples to the spin superfluid, and the nonabelian rotation group reduces to the abelian case with rotations restricted around the $z$ direction. We can identify the accumulated Aharonov-Casher phase as the rotation angle in operator $U$ and $\mathcal{R}$: $\phi_E=2\lambda\int_0^\ell{d\vb*{\ell}'\cdot[\vb{e}_z\times\vb{E}(\ell')]}$, which corresponds to the dimensionless electric flux defined in the main text.

\section{Equations of motion}
The free energy in the easy-plane limit takes the form $\mathcal{F}=\mathcal{F}_\text{B}+\mathcal{F}_\text{J}$, where the bulk energy is given by
 \begin{align}
    \label{eqn: free energy simplified}
    \mathcal{F}_\text{B}
    &=\int_0^L d\ell\left[\frac{A}{2}(\partial_\ell\vb{n})^2+\frac{K}{2}n_z^2 -\vb{D}\cdot(\vb{n}\times\partial_\ell\vb{n})\right]\\&=\int_0^L{d\ell\left[\frac{1}{2A}j_s^2
    +\frac{\tilde{K}-j_s^2/A}{2}n_z^2+\frac{A}{2}\frac{\left(\partial_\ell n_z\right)^2}{1-n_z^2} -\frac{Aa_E^2}{2}\right]}\\  &\approx\int_0^L{d\ell\left[\frac{1}{2A}j_s^2
    +\frac{\tilde{K}}{2}n_z^2 + \text{const.}\right].}
\end{align}
In the second line, we parametrize $\vb{n}$ by its $z$ component $n_z$ and in-plane phase $\varphi$ and use $j_s=-A(\partial_\ell\varphi-a_E)$.
For this, we assumed an in-plane electric field $\vb{E}$ such that $\vb{D}\propto \vb{e}_z$, and introduced the renormalized easy-plane anisotropy $\tilde{K}=K+Aa_E^2$.  
Here, we can see that the DMI induced by in-plane electric field is beneficial for the easy-plane anisotropy, which may provide a strong enough $\tilde{K}$ even with a weak $K$.
In the third line, we assume the Landau criterion is fulfilled, $j_s\ll\sqrt{\tilde{K}{A}}$.
Furthermore, in the strong easy-plane limit, out-of-plane fluctuations are energetically suppressed. 
Hence, the stiffness term $\propto A\left(\partial_\ell n_z\right)^2 $ can be neglected. 
Similarly, an out-of-plane electric field which couples to out-of-plane spin winding is of minor importance.
Therefore, an out-of-plane electric field is not viewed as a parasitic effect and can be safely omitted even when it is of the same order as the in-plane electric field.

The boundary energy at the Josephson junction is given by
 \begin{align}
    \mathcal{F}_\text{J}
    &= -F_\text{J}\vb{n}(0)\cdot\vb{n}(L)\\
    &\approx  -F_\text{J}\cos{(\varphi_L-\varphi_0)},
\end{align}
where we only keep the leading order in $n_z(0)$ and $n_z(L)$. 
The dissipation of the magnetic dynamics can be modeled by Gilbert damping. 
In the long-wavelength limit, the Rayleigh dissipation function is given by~\cite{landauVOL5}
 \begin{align}
    \label{eqn: rayleigh simplified}
    \mathcal{R}
    =\int_0^L{d\ell\frac{s\alpha}{2}(\partial_t\varphi)^2},
\end{align}
 where $\alpha\ll 1$ is the dimensionless Gilbert damping constant.
The bulk equations of motion from the main paper follow directly via
\begin{align}
    {s\partial_t  n_z}&=\{sn_z,{\mathcal F}_\text{B}\} - \frac{\delta{\cal R}}{\delta \partial_t\varphi}, \\
     {\partial_t \varphi}&=\{\varphi,{\mathcal F}_\text{B}\},
\end{align}
where the Poisson bracket $\{\varphi(\ell), s n_z(\ell')\}=\delta(\ell-\ell')$ has to be applied, where $s$ is the linear spin density.

The boundary conditions can be derived by varying the total free energy $\mathcal{F}\left[\varphi,n_z\right]$ to obtain:
\begin{equation}
\begin{split}
   0=\delta\mathcal{F}  &= \int_0^L{d\ell\left[\delta\varphi(\partial_\ell j_s)+\delta n_z(K n_z)\right]} \\
    &+ \delta\varphi_L[-j_s|_{\ell=L}+F_\text{J}\sin{(\varphi_L-\varphi_0)}]  \\ 
   & - \delta\varphi_0[-j_s|_{\ell=0}+F_\text{J}\sin{(\varphi_L-\varphi_0)}],
\end{split} 
\end{equation}
where the first term gives the stationary state for the bulk and the last two terms give the boundary conditions resulting in the continuity of spin current at the weak link.

\section{Electric field generation}
Even in the absence of an external electric field, an electric field $\vb{E}$ can be generated by a nonzero spin current. To see this, we notice that the DMI term from Eq.~\eqref{eqn: DMI} can be rewritten as 
\begin{align}
{\mathcal F}_\text{DMI}= -\int \mathrm{d} V\vb{P}\cdot \vb{E}.
\end{align}
This allows the identification of the electric polarization within the ring in leading order in $\lambda$~\cite{hoshoPRL2005,maximPRL2006} 
\begin{align}
    \vb{P}=\frac{2\lambda}{A_\text{cs}} A\,\vb{e}_\ell \times \left(\vb{n}\times\partial_\ell \vb{n}\right)\approx  -\frac{2\lambda }{A_\text{cs}}\,j_s\vb{e}_n,
    \label{eqn: polarization}
\end{align}
where $A_{cs}$ is the cross section of the ring. 
In the second step, we used the easy-plane limit and introduced the spin current $j_s$. 
The coupling between the electric field and the gradient of the order parameter can be understood as the flexoelectric effect~\cite{liuJAP2012} well studied in the field of liquid crystals~\cite{garbovskiyLCR2025,daoPRL2023}
~\footnote{Based on symmetry, the electric dipole $\vb{P}$ can acquire two distinct contributions: an antisymmetric one from DMI (as in Eq.~\eqref{eqn: polarization}) and a symmetric one of the form $\propto \partial_\ell \left[\vb{n}(\vb{e}_\ell \cdot \vb{n}) \right]$.
While the latter can arise from nonuniform local anisotropy, it vanishes in our system as we assume translational invariance.}. 

To estimate the electric field $\vb{E}$ generated by the polarization and the additional electric flux $\phi_E$ through the ring, we ignore the curvature of the ring ($L \gg w,h$) and assume the cross section to be an ellipse with a major axis (along $\vb{e}_n$) of length $w$ and a minor axis (along $\vb{e}_z$) of length $h$ for simplicity.
Then, using the ellipsoidal symmetry, a uniform polarization $\vb{P}$ leads to a uniform electric field inside the ring
\begin{align}
    \vb{E}=-\vb{N}\vb{P}=-4\pi \frac{h}{w+h}  \vb{P}, \label{eqn:electricfield}
\end{align}
where $\vb{N}$ is the depolarization tensor. In the second step, we used that $\vb{P}\propto \vb{e}_n$, see Eq.~\eqref{eqn: polarization}, and that $\vb{N}\vb{e}_n=N_n \vb{e}_n$ with $N_n=4\pi \frac{h}{w+h}$~\cite{osbornPR1945}. 
Using Eqs.~\eqref{eqn: polarization}-\eqref{eqn:electricfield}, the corresponding electric flux $\phi_E=2\lambda \vb{E}\cdot \vb{e}_n L$ can then be expressed as $\phi_E=L_g j_s$, where in analogy to the magnetic case, we define the geometric inductance
\begin{equation}
    L_\text{g} = 4 \lambda^2 \frac{L}{A_\text{cs}}N_n.
\end{equation}
The energy stored in the electric field created by the spin current is given by
\begin{align}
    {\mathcal F}_E=\int dV\frac{\vb{E}^2}{8\pi}=-\int dV\frac{\vb{E}\cdot{\vb{P}}}{2}=\frac{1}{2} L_\text{g} j_s^2,
\end{align}
where the second integral is nonzero only inside the ring of volume $A_\text{cs} L$.  

However, the main contribution of the bulk energy still stems from the kinetic (or exchange) energy
\begin{align}
    {\mathcal F}_\text{X}= \int_0^L d\ell\frac{A}{2}\left(\partial_\ell\varphi-a_E\right)^2=\frac{1}{2} L_\text{k} j_s^2,
\end{align}
where the kinetic inductance takes the form 
\begin{align}
    L_\text{k}&=\frac{L}{A}.
\end{align}
If the 1D exchange stiffness scales like $A = A_\text{B} A_{cs}$ with $A_\text{B}$ being the 3D bulk value, we obtain the ratio
\begin{align}
    \frac{L_\text{g}}{L_\text{k}}\simeq 4\lambda^2 N_n A_\text{B}.
\end{align}
In general, $\lambda^2 A_B \ll 1$ and $0<N_n<4\pi$, therefore the geometric inductance can be neglected.
This is in contrast to the rf SQUID, where for the dimensions $w$ and $h$ large compared to the London penetration depth, the magnetic energy dominates~\cite{landauVOL9,jacksonWiley1999}.

\normalem
\bibliography{bib_rf_V11}